\documentclass[preprint,showpacs,aps,prc,groupedaddress,floatfix]{revtex4}
\usepackage{epsfig}

\begin{document}

\title{Strong pickup-channel coupling effects in proton scattering:\\ the case of p +
$^{10}$Be}
\author{R. S. Mackintosh}
\email{r.mackintosh@open.ac.uk} \affiliation{Department of Physics
and Astronomy, The Open University, Milton Keynes, MK7 6AA, UK}
\author{N. Keeley}
\email{nkeeley@cea.fr}
\altaffiliation[Permanent address: ]{Department of Nuclear Reactions, The Andrzej
So\l tan Institute for Nuclear Studies, Ho\.za 69, PL-00681, Warsaw, Poland}
\affiliation{CEA/DSM/DAPNIA/SPhN Saclay, 91191 Gif-sur-Yvette Cedex, France}

\date{\today}

\begin{abstract}
The dynamic polarization potential (DPP) contribution to the
effective proton-nucleus interaction, that is due to the coupling of
deuteron channels, is evaluated by applying $S_{lj} \rightarrow
V(r)$ inversion to the elastic channel $S$-matrix from coupled
reaction channel calculations of proton elastic scattering. This was
done for protons scattering from $^{10}$Be at 12, 13, 14, 15, and 16
MeV; non-orthogonality corrections were included. We find a
consistent pattern of a repulsive real and an absorptive imaginary
DPP, with the absorption shifted to a larger radius. This is
consistent with what has been found for proton scattering from the
neutron skin nucleus $^8$He. The DPP is not of a form that can be
represented by a renormalization of the bare potential, and has
properties suggesting an underlying non-local process. We conclude
that deuteron channels cannot be omitted from a full theoretical
description of the proton-nucleus interaction (optical potential).

\end{abstract}


\maketitle

\section{INTRODUCTION}
It is many years since coupled reaction channel (CRC) calculations
first suggested that the coupling to deuteron channels makes a
significant contribution to proton scattering, and in fact
significantly improves the fit to certain
data~\cite{Mac73,Mac76,coulter,kosugi}. The development of inverse
scattering techniques, whereby the potential that yields a specific
S-matrix $S_{lj}$ can reliably be calculated~\cite{Kuk04}, has made
it possible to represent the coupling effects as a contribution to
the dynamic polarization potential (DPP). A consistent
finding~\cite{Mac86,cooper} has been that the deuteron channels
contribute a significant \emph{repulsive} component to the
nucleon-nucleus interaction, in addition to the expected additional
absorptive component.

The CRC calculations cited above all omit the non-orthogonality
corrections~\cite{satchler,Tho88}. It has recently become possible
to include these terms and their importance has been verified in
a recent study~\cite{Ska05} of the contribution of the
$^8$He(p,d)$^7$He pickup reaction to p + $^8$He elastic scattering
at a bombarding energy of 15.7 $A$ MeV. Very strong coupling effects
were found. Although the non-orthogonality terms were found to
modify the details of the DPP quite significantly, the qualitative
features were consistent with earlier findings. Nevertheless, $^8$He
is a far from typical nucleus, the $^8$He(p,d)$^7$He reaction having
a combination of almost perfect Q-matching (Q = $-0.36$ MeV) and a
large spectroscopic factor \cite{Ska05,Kee07} leading to the large
DPP that was found. The present work sets out to demonstrate that
the properties of the DPP found for $^8$He are more generally true.
It is a first step in tracing out the systematics of the evolution
of the effect with incident energy, Q-value and spectroscopic
factor.

Since the nucleon-nucleus interaction is fundamental to nuclear
physics, the nature of the pickup DPP should be verified, especially
in view of the following findings~\cite{Ska05}: (i) the form of the
DPP makes it unrealistic to correct a folding model potential with
real and imaginary normalization factors, as is commonly done; (ii)
there is a local emissive region in the imaginary part of the DPP
which suggests that the local potential is representing non-local
effects (the reference here is not to exchange non-locality) that
would not arise in folding models based on a local density
approximation; (iii) the repulsive nature of the real part of the
DPP is at variance with the findings of dispersive optical models in
which that part of the optical potential, often identified as the
DPP, that is added to the Hartree-Fock component, is
attractive~\cite{jm,tcd,wfprs}.

\section{CRC CALCULATIONS FOR p +$^{10}$Be}\label{crc}
We present here a series of CRC calculations evaluating the
contribution of $^{10}$Be(p,d)$^9$Be pickup coupling to proton
scattering for which a set of $^{10}$Be + p elastic
scattering data at incident proton energies of 12, 13, 14, 15 and 16
MeV is available in the literature~\cite{Aut70}. Fitting this data
helps to ensure that the calculations are realistic. The $^{10}$Be
nucleus provides an excellent opportunity for comparison with
$^8$He, as the number of neutrons is unchanged. The addition of two
protons presents us with a case with a much more negative Q-value
for the (p,d) pickup reaction for $^{10}$Be, $-4.6$ MeV, raising the
possibility that the influence of Q-value on the DPP can be studied.

Although there are unfortunately at present no data for the
$^{10}$Be(p,d) pickup, there are data for the $^9$Be(d,p)$^{10}$Be
stripping reaction at several energies which enable the
$^{10}$Be/$^9$Be spectroscopic factor to be fixed empirically. The shell model
calculation of Cohen and Kurath \cite{Coh67} gives a value of C$^2$S
= 2.36, as does that reported in \cite{Tsa05}, although the
empirical value obtained from the systematic DWBA analysis of
\cite{Tsa05} is significantly lower, $1.58 \pm 0.15$. In this study
we have taken the latter value to give a conservative estimate of
the coupling effect of the $^{10}$Be(p,d) pickup reaction.

The CRC calculations were performed with the code
FRESCO~\cite{Tho88} and included the complex remnant term and
non-orthogonality correction. We took the global nucleon potential
for 1p-shell nuclei of~\cite{Wat69} as a starting point for the
entrance channel potentials. The neutron-proton overlap was
calculated using the Reid soft-core potential \cite{Rei68} and
included the small D-state component of the deuteron ground state.
The n+$^{9}$Be binding potential was of Woods-Saxon form, with a
central part of radius $r_0 = 1.25$ fm and diffuseness $a = 0.65$ fm
and a spin-orbit component of the same geometry and fixed depth of 6
MeV, the depth of the central part being adjusted to obtain the
correct binding energy.

For the d + $^9$Be exit channel, continuum discretized coupled
channels (CDCC) calculations, similar to those described
in~\cite{Kee04} but without the (d,p) coupling, were carried out to
fit the appropriate $^9$Be(d,d) data \cite{Szc89,Abr93,Gri71} taking
the deuteron breakup effect explicitly into account. The n,p +
$^9$Be potentials required as input were based on the global
potential of Ref.~\cite{Kon03}. In order that the CDCC calculations
fit the data, concentrating on the forward scattering angles, the
potentials of Ref.~\cite{Kon03} were renormalized as follows: the
real parts were increased by 30-50 \% and the imaginary parts were
decreased by about 20 \% except for the $^9$Be(d,d) data \cite{Szc89}
at the appropriate energy for the 12 MeV $^{10}$Be(p,d)$^9$Be calculation
where an increase of the imaginary part by a factor of two
was necessary, probably because of the rather low equivalent
deuteron energy. The n-p continuum was discretized into bins of
width $\Delta k = 0.1$ fm$^{-1}$ up to $k_{\mathrm{max}} = 0.4$
fm$^{-1}$. The resulting CDCC input parameters were then
incorporated into the $^{10}$Be(p,d)$^9$Be CRC calculations,
yielding a combined CRC/CDCC calculation similar to that carried out
for the $^8$He(p,d)$^7$He reaction \cite{Ska05}. Transfers to the $L=0$
and $L=2$ unbound states of the n--p system were included. Couplings to the
$L=1$ and $L=3$ states were completely omitted as they are found to have
a small effect \cite{Yah82} on deuteron elastic scattering. Indeed, test
calculations in which the breakup couplings in general were omitted
showed that they had minimal effect in this case, at least on the proton
elastic scattering.

The entrance channel potentials were then re-tuned to obtain
the best fit to the $^{10}$Be(p,p) elastic scattering data.
In practice, it was found that a single potential geometry, slightly
modified from that of the initial global optical potential of \cite{Wat69},
was able to provide good fits to all the data with minor adjustments
to the real and imaginary potential depths. The final ``bare'' potential
parameters are given in Table \ref{tab1}, along with the $\chi^2/N$
values obtained from the full calculations. The parameters of the
spin-orbit potential were fixed: $V_{\rm so}= 5.5$ MeV, $r_{\rm so}
=1.14$ fm  and $a_{\rm so}= 0.57$ fm, as there are no polarization
data available. From the $\chi^2/N$ values in
the last column, it can be seen that the fit becomes relatively poor
at 12 MeV; the change seen in the data beyond 130$^\circ$ for this 1
MeV step is either a resonance-like effect or a problem with the
data itself. In neither case would we expect the data to be fitted
within the optical model framework. For the purpose of calculating
the DPP at 12 MeV, it is reasonable to use the potential that fits
the data for $\theta < 130^\circ$ and has the same geometry as used
at higher energies.
\begin{table}
\caption{\label{tab1} ``Bare'' p + $^{10}$Be potential parameters
and $\chi^2/N$ values for the full CRC calculations. The real and
imaginary potentials are of volume and surface Woods-Saxon form,
respectively.}
\begin{ruledtabular}
\begin{tabular}{c c c c c c c c}
Energy (MeV) & $V$ (MeV) & $r_V$ (fm) & $a_V$ (fm) & $W_D$ (MeV) & $r_D$ (fm) & $a_D$ (fm) & $\chi^2/N$ \\
16 & 65.7 & 1.137 & 0.514 & 7.15 & 1.068 & 0.497 & 3.46 \\
15 & 65.7 & 1.137 & 0.514 & 7.15 & 1.068 & 0.497 & 2.38 \\
14 & 65.7 & 1.137 & 0.514 & 6.00 & 1.068 & 0.497 & 1.93 \\
13 & 65.0 & 1.137 & 0.514 & 6.70 & 1.068 & 0.497 & 1.00 \\
12 & 65.0 & 1.137 & 0.514 & 6.00 & 1.068 & 0.497 & 8.52 \\
\end{tabular}
\end{ruledtabular}
\end{table}
\begin{figure}
\caption{\label{fig1}Data for $^{10}$Be(p,p) elastic scattering
\cite{Aut70} compared with the full CRC calculations (full curves)
and the no-coupling calculations (dashed curves). (a) 16 MeV, (b) 15
MeV, (c) 14 MeV, (d) 13 MeV, (e) 12 MeV; plotted as ratio to
Rutherford in each case.}
\begin{center}
\psfig{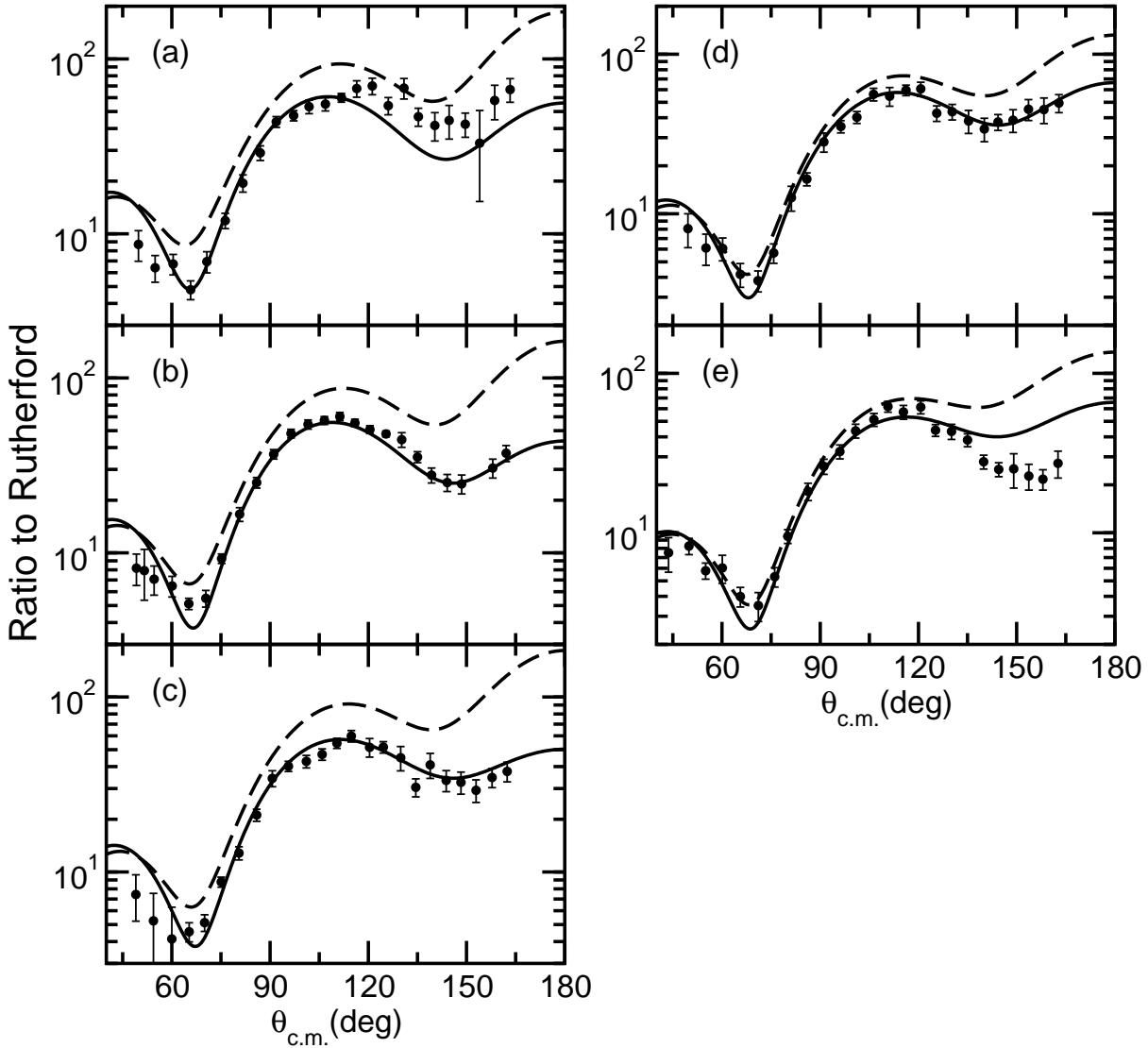}
\end{center}
\end{figure}

The calculations are compared to the data in Figure~\ref{fig1} where
we also plot the ``bare'' no-coupling results. The good agreement
with the data is borne out by the $\chi^2/N$ values given in
Table~\ref{tab1}. With a final total of four adjustable
parameters (real and imaginary potential depths in the entrance channel
and real and imaginary potential normalization parameters in the
exit channel) this is as it should be, although the real and imaginary
potential normalization factors for the $^9$Be(d,d) CDCC
calculations in the exit channels were determined separately against
the appropriate elastic scattering data and then held fixed.
However, the motivation of the current work was to determine the DPP
due to the (p,d) pickup as accurately as possible, hence the need
for the best possible fit to the data.

It is apparent from a comparison of the full and dashed curves in
Fig.~\ref{fig1} that the effect of the (p,d) pickup coupling remains
important for the $^{10}$Be(p,p) elastic scattering at these
energies, although considerably less than was found for the
$^8$He(p,p) elastic scattering. In the next section we describe the
effect upon the DPP of increasing the spectroscopic factor to the
value used in the $^8$He + p scattering analysis.
It is also apparent from Fig.~\ref{fig1} that the coupling effect at 12
and 13 MeV is significantly smaller than at the slightly higher
energies.

\section{CALCULATION OF THE DPP}
We apply $S_{lj} \rightarrow V(r)$ inversion~\cite{Kuk04} to the
diagonal (elastic scattering) part of the $S$-matrix produced in the
CRC calculations described above to yield a potential $V_{\rm
crc}(r)$. The inversion is carried out using the
iterative-perturbative (IP) method~\cite{bu1,andy,Kuk04}. The local
potential found by inversion would, if inserted into an optical
model (single channel) code, precisely reproduce the theoretical
elastic scattering from the CRC calculations. The potential $V_{\rm
crc}(r)$ is, of course, complex and contains a complex spin-orbit
term. The difference $V_{\rm dpp}(r)= V_{\rm crc}(r)- V_{\rm
bare}(r)$, between $V_{\rm crc}(r)$ and the bare potential $V_{\rm
bare}(r)$, is a local and $L$-independent representation of the
contribution to the dynamic polarization part of the proton-nucleus
potential that is generated by the coupling to the pickup channels.

\begin{figure}
\caption{\label{figpot3}Potentials for 14 MeV $^{10}$Be(p,p):
inverted potential (full curve) and bare potential (dashed curve).}
\begin{center}
\psfig{figure=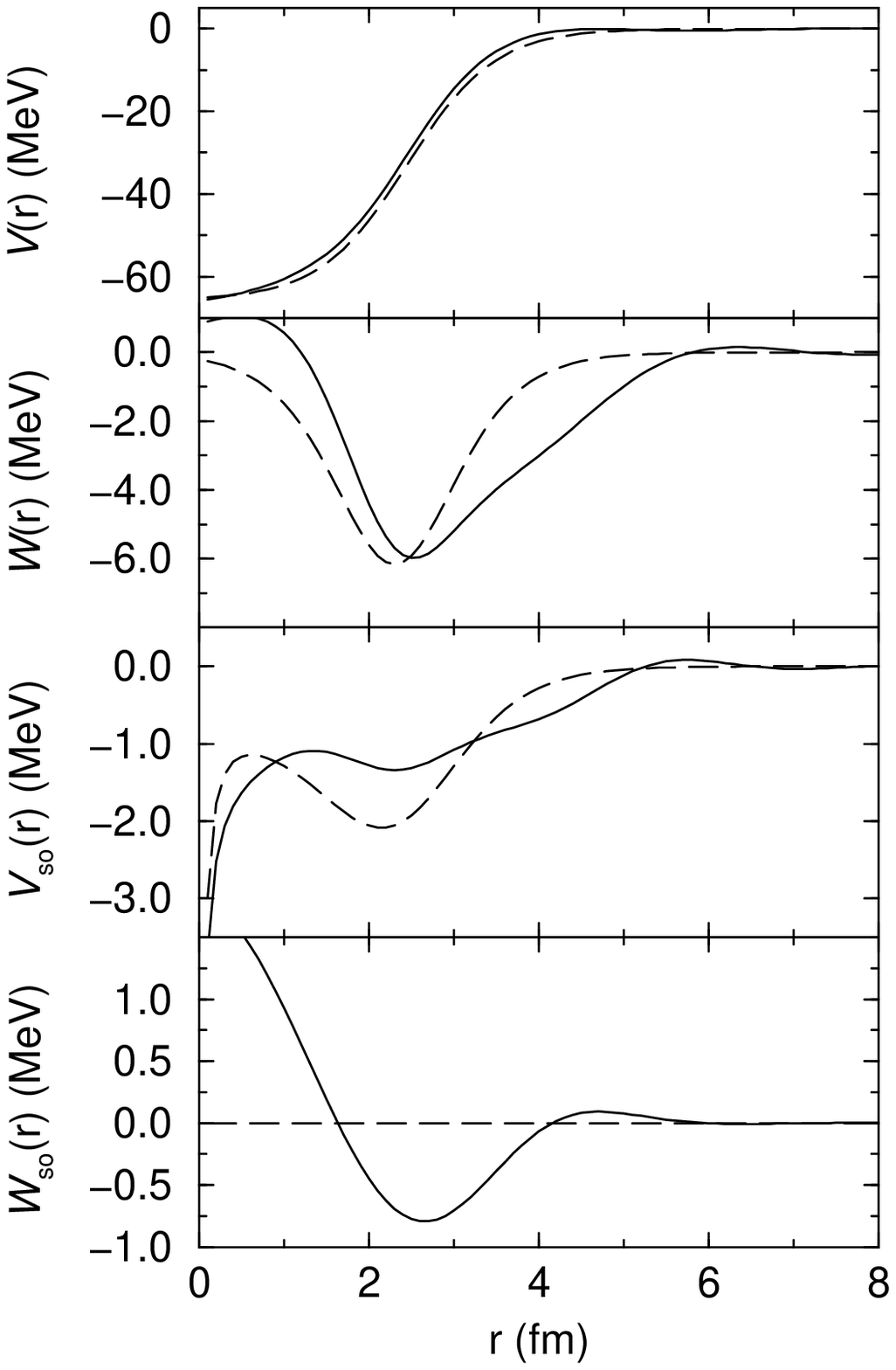,width=10cm,clip=}
\end{center}
\end{figure}

In Figure~\ref{figpot3}
we compare the bare potentials $V_{\rm bare}(r)$ (dashed curve) with
$V_{\rm crc}(r)$ (solid curve) at 14 MeV. The effect is
qualitatively the same at all five energies, although the magnitude
of the effect falls off at the lowest energies, as might be expected
by comparing the 12 MeV fit with those at 14 -- 16 MeV in
Figure~\ref{fig1}. The predominant effect on the real part at each
energy is repulsive and is not negligible, amounting, as quantified
below, to an almost 10 \% effect. The effect on the imaginary part
is quite large, in each case moving the absorptive region
significantly outwards to a larger radius. The modification of the
potentials can certainly not be represented as a multiplicative
factor times the bare potential.

\begin{figure}
\caption{\label{figdpp1}Dynamic polarization potential, DPP, for 14,
15 and 16 MeV $^{10}$Be(p,p) elastic scattering.}
\begin{center}
\psfig{figure=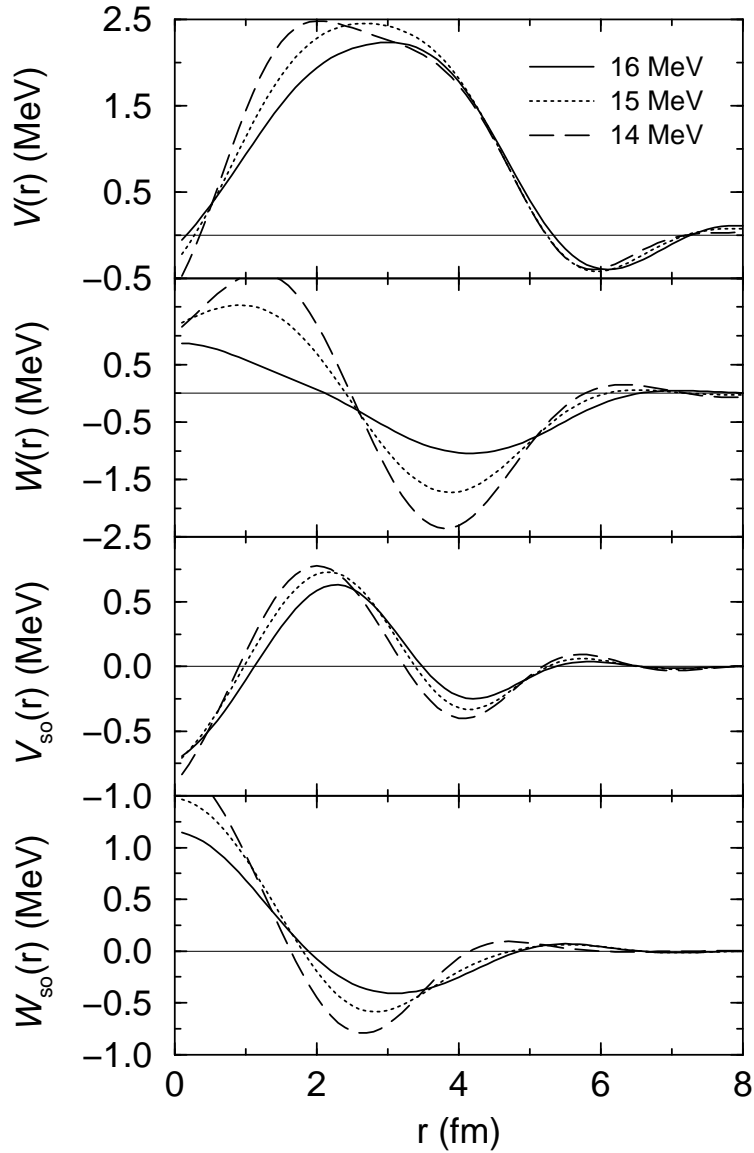,width=10cm,clip=}
\end{center}
\end{figure}

\begin{figure}
\caption{\label{figdpp2}Dynamic polarization potential, DPP, for 12,
13 and 14 MeV $^{10}$Be(p,p) elastic scattering.}
\begin{center}
\psfig{figure=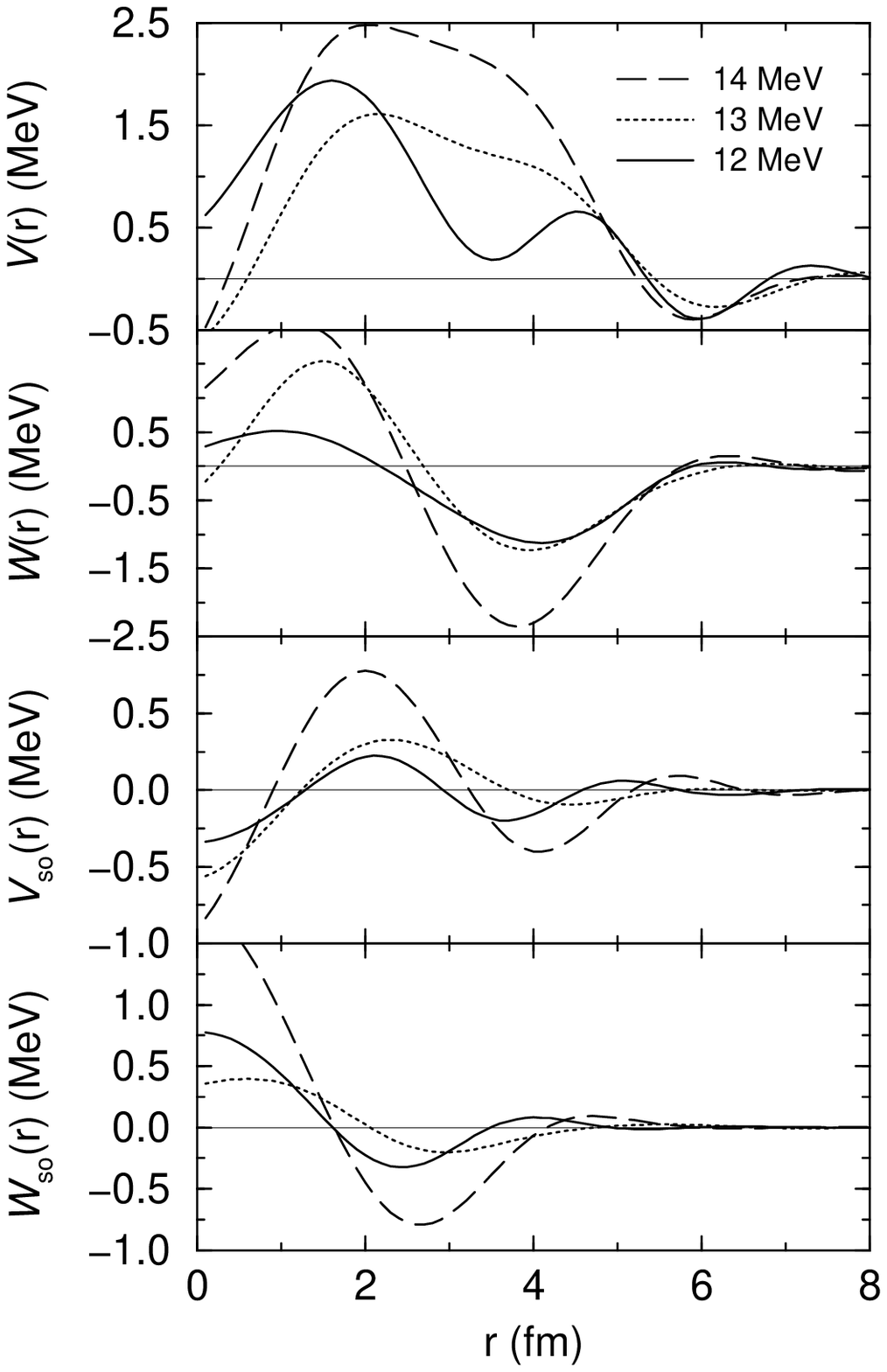,width=10cm,clip=}
\end{center}
\end{figure}

The contributions from the coupled reaction channels are more
clearly seen by examining the DPP itself, $V_{\rm dpp}(r)$,
presented for each energy in Figures~\ref{figdpp1}
and~\ref{figdpp2}; the 14 MeV DPP is given in each to facilitate
comparisons.

The imaginary part of the central DPP is particularly large, showing
a clear emissive region at smaller radii. This does not, of course,
correspond to any overall emissive regions in the potential, nor any
unitarity breaking, but is significant. Such emissive features are
commonly found in DPPs that are generated by coupled channels, and
probably relate to the fact that the DPP, as presented here, is a
local and $L$-independent representation of a term that, in a full
Green's function treatment, would be non-local and $L$-dependent.
This non-locality is quite distinct from that arising from exchange,
and the $L$-dependence referred to is distinct from the parity
dependence arising from heavy particle stripping,
see~\cite{feshbach}.

The pickup contributions can usefully be quantified using the
conventionally defined~\cite{satchler} real and imaginary volume
integrals, $J_{\rm R}$ and $J_{\rm I}$ and rms radii, $R_{\rm R}=
(\sqrt{\langle r^2 \rangle})_{\rm R}$ and $R_{\rm I}=(\sqrt{\langle
r^2 \rangle})_{\rm I}$. The three most conspicuous changes are a
decrease in $J_{\rm R}$ (a repulsive effect) an increase in $J_{\rm
I}$ (absorption) and an increase in $R_{\rm I}$, representing the
net shift of the absorptive part of the potential to a larger
radius.
Table~\ref{tab2}
presents the changes induced by coupling to these three
quantities for each energy. For comparison, the bare potential at 15
and 16 MeV has $J_{\rm R}= 576.1 $ (MeV fm$^3$) and $J_{\rm I}=
108.8$ (MeV fm$^3$), so the changes are of the order of 10\% for the
real part and 50 \% for the imaginary part.
\begin{table}[hbt]
\begin{center}
\begin{tabular}{|c|r|r|r|} \hline
$E_{\rm lab}$ (MeV)  & $\Delta
J_{\rm R}$ (MeV fm$^3$) & $\Delta J_{\rm I} $ (MeV fm$^3$) & $\Delta R_{\rm I}$ (fm)\\
\hline  16  & $-66.9$ & 57.3 & 0.623 \\
 15  & $-68.8$ & 67.9 & 0.626 \\
 14  & $-67.9$ &77.2 &0.694 \\
 13  & $-44.9$& 47.2&0.699\\
 12  & $-24.4$ & 50.1 & 0.613\\
\hline
\end{tabular}\caption{\label{tab2}The changes
in selected characteristics of the proton-nucleus interactions
induced by pickup coupling.}
\end{center}
\end{table}

Apart from the 12 MeV case, the results are rather consistent. It
should be noted that all the quantities given in this table are
relatively small differences between pairs of quantities, one of each
pair being subject to uncertainties that are hard to evaluate
precisely. At lower energies, the number of active partial waves
falls, and the linear equations upon which the IP inversion
procedure centers become less definitely over-determined. As a
result the IP inversion procedure yields potentials that may be less
well-determined; in particular they may have small wiggles in the
surface that can contribute disproportionately to the volume
integrals and rms radii. For this reason, it is the qualitative
properties of the DPP rather than point-by-point values that should
be considered well-determined at 12 and 13 MeV.

Nevertheless, the general properties of the DPP can be considered to
be well established. In particular, the smaller magnitude of $\Delta
J_{\rm I}$ at 13 and especially 12 MeV appears to be a dynamic
effect and not an artifact of the inversion. This is supported by
the fact that the difference between the solid and dashed lines in
the 12 MeV and 13 MeV cases in Figure~\ref{fig1}, over the mid-angle
range, is substantially less (noting that these are logarithmic
plots) than for the other three figures corresponding to 14 --- 16
MeV. These figures suggest that the CRC model itself entails a
smaller DPP as the energy falls below 14 MeV; the small changes in
the bare potentials given in Table~\ref{tab1} are not expected to
have a dramatic effect on the DPPs.  Note that the outward shift in
the imaginary potentials is as large at 12 MeV as it is at the
higher energies.

The general properties of the DPP are similar to those
found~\cite{Ska05} in the scattering of 15.7 MeV protons on $^8$He.
In that case $\Delta J_{\rm R} =-51.2$ (MeV fm$^3$), $\Delta J_{\rm
I} = 261.41 $ (MeV fm$^3$) and  (not reported in~\cite{Ska05})
$\Delta R_{\rm I} = 0.827$ fm. The much larger value of $\Delta
J_{\rm I}$ in that case might be attributed to the near zero Q-value
and much larger spectroscopic factor (2.9, compared to 1.58 for the
$^{10}$Be case). In addition, the bare proton potential in the
$^8$He case had a very small imaginary part, $J_{\rm I}= 33$ (MeV
fm$^3$), the deuteron channel contributing a large proportion of the
reaction cross section. In a model calculation for p + $^{10}$Be at
13 MeV, in which we increased the spectroscopic factor from 1.58 to
2.9, the $^8$He value, we found the following: $\Delta J_{\rm R}
=-73.0$ (MeV fm$^3$), $\Delta J_{\rm I} = 100.2 $ (MeV fm$^3$) and
$\Delta R_{\rm I} = 1.004$ fm. In Section II we noted that we had
chosen the spectroscopic factor to give a conservative estimate of
pickup coupling contributions.

All the above results were obtained using bare potentials that had
the same geometry, but which had minor adjustments to the depths to
give the best CRC fits to the elastic scattering. The optimum
parameters for such potentials can not be determined uniquely with
data of the range and precision seen in Figure~\ref{fig1}. The
question then arises as to how the results might depend upon the
bare potential and we performed a small number of calculations with
different potentials. The general result is that the qualitative
results do not depend upon the specific potential, but the specific
magnitudes of the effects do depend somewhat upon the details of the
bare potential. Further explorations of this matter will hopefully
lead to an understanding of such questions as to why the real DPP is
repulsive. See Reference~\cite{andy} for an earlier discussion of
this but in the context of zero-range CRC without non-orthogonality
corrections.

\section{CONCLUSIONS: IMPLICATIONS FOR UNDERSTANDING NUCLEON-NUCLEUS
INTERACTIONS}\label{conclusions} We have presented a local and
$L$-independent potential, generated by the coupling to deuteron
channels, for the case of protons scattering from $^{10}$Be.  The
overall properties of the DPP were qualitatively similar to what was
found in the case of scattering from neutron skin nucleus
$^8$He~\cite{Ska05}, although the contribution to the overall
absorptive potential was less than for $^8$He. We therefore propose
that the effects found there were not peculiar to proton scattering
from $^8$He, and the overall repulsive/absorptive effect is an
example of a more general phenomenon. As in all cases studied
previously, prior to the inclusion of non-orthogonality corrections,
the real part had an overall repulsive character. A small imaginary
spin-orbit interaction was generated.

The complex DPP that we found was not at all of a form that could be
represented by  renormalizing a folding model potential. We
therefore conclude that the determination of such normalization
factors is not a satisfactory way of evaluating folding models. It
would seem preferable to use model-independent fitting to determine
an additive component to a folding model potential. Ideally, this
would fully exploit the information content of the data to yield
empirical DPPs that could be compared with those calculated in
studies such as this.

Some basic problems must be acknowledged. There exists no numerical
implementation of a fully rigorous reaction theory, certainly not
one based on realistic nucleon-nucleon interactions. As a result
there are inevitable uncertainties in the interpretation of our
results. For example, the deuteron channel states are not orthogonal
to particle-hole states. Such particle-hole states enter into any
local density model devised to handle realistic nucleon-nucleon
interactions. Moreover, strictly speaking, the concept of the DPP
corresponding to specific channels rests on the orthogonality of the
coupled channels~\cite{coll}. There is therefore an unresolved
double counting problem. Nevertheless, the CRC calculation does
provide a representation of processes that would not be present in a
local-density folding model. This is suggested, for example, by the
form of the imaginary DPP with its emissive feature at small $r$.
The underlying non-locality that this suggests probably corresponds
to the fact that the deuteron in the intermediate state is
propagating in a potential with a strong gradient, something
local-density models do not encompass.

The nature of the DPP presented here does not fit naturally into the
generally accepted understanding of nucleon-nucleus scattering and
it is natural to ask for supporting evidence.  The present rather
good fits to the elastic scattering data for protons on $^{10}$Be
can equally well be reproduced by a local single-channel optical
model. However, other cases do exist~\cite{Mac76}, albeit studied
prior to the inclusion of non-orthogonality terms, in which data
could be fitted when the coupling to pickup channels was included
that persistently resisted fitting with conventional optical
potentials (at least smooth, $L$-independent potentials);  we intend
to pursue such cases in the future. Further in support of our
conclusions, we note that the general CRC formalism used here, as
applied to the elastic scattering of heavy ions, has very
successfully explained such phenomena as the threshold anomaly in a
range of systems, e.g.\ \cite{tho89,kee96}. We therefore feel
confident that we have demonstrated that \emph{the coupling to
deuteron channels must be included in a full account of nucleon
scattering from nuclei.} It is possible that local density models
include some of the effect \emph{in an average way}, in which case
the challenge will be to relate the specific pickup channels for
specific nuclei to irregularities in the $N$ and $Z$ dependence in
elastic scattering, something that is essential before the
interaction between nucleons and nuclei can be said to be
understood.

\begin{acknowledgments}
N.K. gratefully acknowledges the receipt of a Marie Curie
Intra-European Fellowship from the European Commission, contract
No.\ MEIF-CT-2005-010158.
\end{acknowledgments}

\end{document}